\documentclass[aps,preprint,prd,showpacs,nofootinbib]{revtex4-1}

\usepackage{epsf}

\usepackage{bm}% bold math
\expandafter\let\csname equation*\endcsname\relax
\expandafter\let\csname endequation*\endcsname\relax
\usepackage{amsmath,amssymb}
\usepackage{amsfonts}
\usepackage{subfigure}
\usepackage{graphicx}
\usepackage{natbib}

\newcommand{\mn}{{\mu\nu}}

\def\be{\begin{equation}}
\def\ee{\end{equation}}
\def\bea{\begin{eqnarray}}
\def\eea{\end{eqnarray}}

\newcommand{\bmat}{\left(\begin{array}}
\newcommand{\emat}{\end{array}\right)}

\newcommand{\bit}{\begin{itemize}}
\newcommand{\eit}{\end{itemize}}
\newcommand{\bnu}{\begin{enumerate}}
\newcommand{\enu}{\end{enumerate}}

\newcommand{\ba}{\begin{align}}
\def\mn{{\mu\nu}}

\begin{document}

\title{$f(R)$ gravity constraints from gravitational waves}

\author{Jaakko Vainio and Iiro Vilja}
\affiliation{Department of Physics and Astronomy, Turku Center for Quantum Physics, University of Turku, FIN-20014, Finland}
\email{jaakko.vainio@utu.fi, vilja@utu.fi}

\begin{abstract}
The recent LIGO observation sparked interest in the field of gravitational wave signals. Besides the gravitational wave observation the LIGO collaboration used the inspiraling black hole pair to constrain the graviton mass. Unlike general relativity, $f(R)$ theories have a characteristic non-zero mass graviton. We apply the constraint on the graviton mass to viable $f(R)$ models to find the effects on model parameters. We find it possible to constrain the parameter space with the gravity wave based observations. We make a case study for the popular Hu-Sawicki model and find a parameter bracket. The result generalizes to other $f(R)$ theories and can be used to contain the parameter space.
\end{abstract}

\maketitle

\section{Introduction}
The recent observation of gravitational waves \cite{PhysRevLett.116.061102} confirmed the hundred years ago predicted gravitational waves. In the early years of general relativity different models for gravitation were considered as well. For a long time these alternatives to general relativity were little more than a curiosity as the observations of that time did not call for anything else. Many of these modified theories of gravity were ruled out for theoretical reasons but others remained viable.

Cosmic microwave background \cite{Bunn1996} and supernova observations \cite{perlmutter99, riess98} lead to the discovery of accelerating expansion of the Universe. It can be argued that this discovery and a growing body of observations lead to a renaissance in cosmology. The accelerated expansion could be explained with the cosmological constant, but there are some fundamental problems with the cosmological constant \cite{Weinberg:1988cp} and the $\Lambda$CDM or concordance model \cite{Wang:1999fa}. Therefore, the modified gravity theories which received little interest for decades have become relevant once again.

The $f(R)$ theories (see {\it e.g.} \cite{Sotiriou:2008rp, DeFelice:2010aj} for reviews), or fourth order theories, which generalize the Einstein Hilbert Langrangian to be a function of the curvature scalar, have received considerable attention in the 21st century. In \cite{Vollick2003} it was shown that the accelerating expansion could be explained with a $f(R)$ modification. Since then, more viable models have been proposed ({\it e.g.} \cite{Hu2007, Starobinsky2007, Appleby:2007vb, Tsujikawa2007, Cognola:2007zu, Nojiri:2007as}).

In standard general relativity the graviton, which mediates the gravitational force, has a zero mass. In order to give graviton a mass some generalization is needed, namely taking a set background metric \cite{Visser:1997hd}. General relativity is a unique theory given a certain set of postulates\footnote{A metric theory with field equations of linear second order in derivatives, satisfies the Newtonian weak field limit and does not depend on any prior geometry.} and the path of least change is fixing the background metric.

It is possible to add a term to the Einstein Hilbert action causing a massive graviton \cite{Visser:1997hd, Sutton:2001yj}. There are a number of different terms that produce a massive graviton but most of these fail to reach the correct Newtonian limit \cite{Boulware:1973my, Ford:1980up}. However, while in general relativity the graviton naturally has a zero mass, this is not the case for $f(R)$ gravity \cite{Sotiriou:2008rp}.

In $f(R)$ gravity the graviton has {\it a priori} a non-zero mass. As the $f(R)$ theories are explicitly higher order theories, this in not in contradiction with the demands of constructing a massive graviton for general relativity. The higher order contribution in the field equations adds up to effective graviton mass term. This link between graviton mass and model dependence can be converted into boundaries for viable $f(R)$ models.

Solar system observations have set several bounds on the mass of the graviton. As the dynamics of the solar system are found to follow general relativity extremely closely, these bounds are rather stringent. If the Newtonian potential is modified with the graviton mass, the Kepler laws produce a limit for the Compton wavelength of the graviton \cite{PhysRevLett.61.1159, Will:1997bb}. As the Compton wavelength is related to the mass \cite{TheLIGOScientific:2016src} via $\lambda_g=h/m_g c$ this translates to a bound on graviton mass.

Inspiraling binaries are a known source of gravitational waves and a possibility to commit graviton mass measurements \cite{Will:2005va, PhysRevLett.61.1159, Will:1997bb}. Before the LIGO experiments the graviton mass has been bounded by binary pulsars \cite{Sutton:2001yj} instead of a pair of black holes. Assuming a non-zero mass $m_g$ graviton would cause the gravitational potential to be of the Yukawa form $r^{-1}e^{-m_grc/h}$. The exponential dependence would cause a cut-off of the gravitational interaction at large distances, namely larger than the Compton wavelength. Such a cut-off has not been observed in the solar system \cite{PhysRevLett.61.1159} or galaxy clusters \cite{PhysRevD.9.1119}. Therefore, these observations set an upper limit for the mass of the graviton $m_g$.

The galaxy cluster limits for the graviton mass are the most stringent ones with $m_gc^2<2\times 10^{-29}\text{eV}$ \cite{PhysRevD.9.1119}, but are model dependent regarding {\it e.g.} dark matter assumptions. These are not directly applicable to $f(R)$ theories as they modify the effects and need of dark matter \cite{Boehmer:2007kx, Nojiri:2008nt, Bertolami:2007vu, Lobo:2008sg}. For the time being, the best model independent bounds for the graviton mass are those from the recent LIGO observations $m_gc^2<1.2\times 10^{-22}\text{eV}$ \cite{TheLIGOScientific:2016src}. If a super massive black hole binary is detected in the future, it could introduce a several orders of magnitude more stringent limit \cite{Will:2005va}. The gravitational wave based bounds arise from the dynamics of gravitation and as such are model-independent.

In the following we will examine the naturally occurring graviton mass in the $f(R)$. There have been several studies into constraining $f(R)$ theories with both theoretical and observational means ({\it e.g.} \cite{Faulkner2006, Cataneo:2014kaa, DeFelice:2010aj, Turyshev:2008dr, PhysRevLett.70.2220, PhysRevD.74.104017,Dolgov2003}). With the recent LIGO upper limit on the graviton mass we can further constrain the model parameters of viable $f(R)$ theories such as the Hu-Sawicki model \cite{Hu2007}.

\section{Equations of motion}
In the following we derive the equations of motion describing gravitational waves and graviton mass arising from the $f(R)$ contribution. We examine a $f(R)$ modified gravitational action\footnote{The signature of the metric is $-,+,+,+$, the Riemann curvature tensor is $R^\alpha_{\beta\mu\nu}=\partial_\mu\Gamma^{\alpha}_{\beta\nu}-\partial_\nu\Gamma^{\alpha}_{\beta\mu}+\Gamma^\alpha_{\kappa\mu}\Gamma^{\kappa}_{\beta\nu}-\Gamma^\alpha_{\kappa\nu}\Gamma^\kappa_{\beta\mu}$ and the Ricci tensor is $R_{\mu\nu}=R^\alpha_{\mu\alpha\nu}$.}
\be
\mathcal A =\frac 1{2\chi}\int d^4x\sqrt{-g}\Big(f(R)+2\chi \mathcal L_m\Big),
\ee
where $\chi=\frac{8\pi G}{c^4}$ is the coupling of gravitational equations. The latter term $\mathcal L_m$ is the minimally coupled matter Lagrangian. Following standard metric variational techniques we find the field equations and the trace equation
\ba
f'(R)R_{\mu\nu}-\frac 12f(R)g_{\mu\nu}-\nabla_\mu\nabla_\nu f'(R)+g_{\mu\nu}\square f'(R)&=\chi T_{\mu\nu} \label{eqmo}\\
3\square f'(R) + f'(R)R-2f(R)&=\chi T \label{eqmot},
\end{align}
where $T_{\mu\nu}=-\frac 2{\sqrt{-g}}\frac{\delta\sqrt{-g}\mathcal L_m}{\delta g^{\mu\nu}}$ is the energy-momentum tensor and $T=T^\alpha_\alpha$. The prime is used to denote the derivatives with respect to $R$. We study the linear perturbations $h_\mn$ and write
\be
g_\mn=\tilde g_\mn+h_\mn,
\ee
where $\tilde g_\mn$ is the background metric. In general we use tilde to denote the quantities calculated with the background metric. The Ricci tensor and scalar can be expanded around the background as
\ba
R_\mn &\simeq \tilde R_\mn+\delta R_\mn+\mathcal O(h^2), \\
R &\simeq \tilde R+\delta R+\mathcal O(h^2).
\end{align}
As the first derivative of $f(R)$ appears in the equations of motion, we need an expansion for this function as well, {\it i.e.} $f'(R)\simeq f'(\tilde R)+f''(\tilde R)\delta R+\mathcal O(4)$. This expansion can be plugged into \eqref{eqmot} for
\be
f''(\tilde R)(3\square\delta R+\tilde R\delta R)-f'(\tilde R)\delta R=0 \label{deqmo}.
\ee
As we are primarily interested in the propagation of gravitational waves in empty space, we set $T_\mn=0$. The variations of the Ricci tensor and scalar can be written in terms of the metric perturbation $h_\mn$ ({\it e.g.} \cite{weinberg1972})
\ba
\delta R_\mn&=\frac 12\Big(\nabla_\mu\nabla_\nu h-\nabla_\mu\nabla^\lambda h_{\lambda\nu}-\nabla_\nu\nabla^\lambda h_{\mu\lambda}+\square h_\mn\Big), \\
\delta R&=\delta(g^\mn R_\mn)=\square h-\nabla^\mu\nabla^\nu h_\mn-\tilde R_\mn h^\mn. \label{deltar}
\end{align}

As the case is gauge invariant we fix the gauge to be the harmonic gauge with
\be
\nabla_\mu h^\mu_\lambda=\frac 12\nabla_\lambda h,
\ee
which further implies $\nabla^\mu\nabla^\nu h_\mn=\frac 12\square h$. 
%For general relativity, $f(R)=R$ we would have very simple equation of motion in empty space
%\be
%\square h_\mn=0. \label{grwave}
%\ee
%In the harmonic gauge, inserting \eqref{deltar} into \eqref{deqmo} produces
%\be
%\frac 32f''(\tilde R)\square^2 h+\Big(\frac 54\tilde R f''(\tilde R)-\frac{f'(\tilde R)}2\Big)\square h+\Big(\frac{\tilde R^2 f''(\tilde R)}{4}-\frac{f(\tilde R)\tilde R}4\Big)h=0.
%\ee
%\be
%m_g^2=\frac{f'(\tilde R)-\tilde R f''(\tilde R)}{3 f''(\tilde R)}=\frac{f'(\tilde R)}{3f''(\tilde R)}-\frac {\tilde R}3.
%\ee

For a viable $f(R)$ theory to have a de Sitter solution, the background equations of \eqref{eqmo} and \eqref{eqmot} for empty space, $f'(\tilde R)\tilde R=2f(\tilde R)$ and $\tilde R_\mn=g_\mn\frac{f( \tilde R)}{2f'(\tilde R)}$, must hold. Using these equalities and the harmonic gauge we find
\be
3f''(\tilde R)\square^2 h-\Big(\frac{f(\tilde R) f''(\tilde R)}{f'(\tilde R)}+f'(\tilde R)\Big)\square h+\Big(f(\tilde R)-\frac{2f^2(\tilde R)f''(\tilde R)}{f'^2(\tilde R)}\Big)h=0.
\ee
The graviton dispersion relation $k^2=-m_g^2$ reveals that the plane wave solution $h\sim e^{ik \cdot x}$ fulfills $\square h=m_g^2h$. Therefore, we can write

%The empty space equations, including the graviton \cite{Sutton:2001yj}, are $(\square-m_g^2) h_\mn=0$, which lead to $k^2=-m_g^2$ and further $\square h=m_g^2h$, for a plane wave $h\sim e^{ik_\lambda x^\lambda}$. Here and for the rest of the paper we have assumed natural units. Therefore, we can write
\be
3f''(\tilde R)m^4_g-\Big(\frac{f(\tilde R) f''(\tilde R)}{f'(\tilde R)}+f'(\tilde R)\Big)m^2_g+\Big(f(\tilde R)-\frac{2f^2(\tilde R)f''(\tilde R)}{f'^2(\tilde R)}\Big)=0,
\ee
for non-zero perturbations. Thus we obtain two solutions for $m_g^2$
\ba
m_1^2&=\frac{f'^2(\tilde R)-2f(\tilde R)f''(\tilde R)}{3f'(\tilde R)f''(\tilde R)}, \label{eq1}\\
m_2^2&=\frac 12\tilde R, \label{eq2}
\end{align}
which tell us the perturbations of the metric can be written as a linear combination
\be
h_\mn=h^{(1)}_\mn e^{ik^{(1)}\cdot x}+h^{(2)}_\mn e^{ik^{(2)} \cdot x},
\ee
where the quantities $h^{(i)}_\mn$ and $k^{(i)}_\mu$ are the metric perturbation and four-momentum related to the corresponding solution $m_i$.

We have found two physically viable solutions for a non-zero graviton mass. The first solution \eqref{eq1} resembles the stability criterion of \cite{Faraoni:2005vk,Faraoni:2007yn}. Basically this criterion tells us, that the square of the graviton mass must not be negative. The mass is often derived with the well-known $f(R)$ theory scalar-tensor theory equivalence \cite{Yang:2011cp, Capozziello:2008rq, Corda:2010zza}. This solution is not available when $f''(R)=0$, such as in the case of GR.

The second solution \eqref{eq2} does not depend on $f''(R)$ and would hold even for GR. This solution is related to having $\delta R=0$ in \eqref{deqmo}. In the case of empty space GR we would have $\tilde R=0$ and $m_2=0$ as is to be expected.  Clearly, there is a well-behaved GR limit, $f''(R)\to 0$, for the second solution. Since for this solution $\delta R=0$, in the situation $\tilde R=0$ the perturbation of the metric would be simply
\be
\delta R\sim h^{(1)}_\mn e^{ik^{(1)} \cdot x}
\ee
and only the scalar modes would manifest. Therefore, $m_2$ solutions do not effect scalar perturbations while the tensor perturbations are affected by both the solutions.

For the first solution, the GR limit is problematic as it diverges as $f''(R)\to 0$. This reveals an interesting fact that even though $f(R)$ models have to closely resemble GR, it cannot be infinitely close. This can be compared to the result of the forbidden Higuchi mass range of the graviton \cite{HIGUCHI1987397, HIGUCHI1989745, Dilkes:2001av}. We also notice that the second solution is extremely small with $m_2\sim\sqrt \Lambda$, which easily passes all constraints on graviton mass. Therefore, our attention concentrates on the first solution, which can be constrained. It is unknown which mass state of gravitons inspiraling black holes would emit. Mergers in $f(R)$ gravity would need to be studied further to be able to distinguish between these two. To our knowledge, such studies have not yet been conducted. 

Another, often overlooked, fact is that for GR with $\Lambda$, $f(R)=R+\Lambda$, we would have a non-zero graviton mass, $m_2^2=2\Lambda$. This is due to relaxing the assumptions of GR \cite{Visser:1997hd}. Even though this is mathematically clear, the physical consequences are debatable, see {\it e.g.} \cite{Gazeau:2006uy} and references therein for discussion. 

For the case of $f(R)$ gravity, there is the extra scalar degree of freedom like with the cosmological constant. A massive graviton always implies extra degrees of freedom. Due to the added degrees of freedom, the gravitational waves with $\Lambda$ or $f(R)$ are different to those caused by plain GR. However, this does not affect the relation to observations.

The LIGO observations provide a lower limit for the Compton wavelength of the graviton \cite{PhysRevLett.116.061102}. A finite Compton wavelength in general, would translate to a massive theory and therefore, extra degrees of freedom. The measurements detect perturbations of the metric, $h_\mn$, which can be written as a linear combination of the modes associated with masses \eqref{eq1} and \eqref{eq2}. It is not known, what is the ratio of these two modes caused by the black holes but the total contribution is constrained.

In the following, we shall take a closer look at specific models and use the Hu-Sawicki model as a case study to demonstrate the procedure.

\section{Viable $f(R)$ models and graviton mass constraints}
There have been numerous studies to constrain viable $f(R)$ models \cite{Cataneo:2014kaa, Hu2007, Jain:2012tn}. The most stringent bound with $\tilde R$ is
\be
|f'(\tilde R)-1|<4\times 10^{-7},\label{fr0}
\ee
constraining the parameters of the $f(R)$ function.  Here, and for the rest of the paper we assume natural units. With the graviton mass we can find another bound for these parameters.

The popular Hu-Sawicki model \cite{Hu2007} is constructed to evade the solar system tests and produce the observed late-time cosmology. A truly viable model needs to fulfil the high curvature regime constraints as well as provide the accelerated expansion of the Universe, which appears at low curvature regimes. The Hu-Sawicki model is of the form
\be
f(R)=R-\mu R_c \frac{\Big(\frac R{R_c}\Big)^{2n}}{b \Big(\frac R{R_c}\Big)^{2n}+1},
\ee
with $\mu$, $R_c$, $b$ positive constants and $n\in\mathbb N$. Inserting this into the de Sitter criterion, $\tilde Rf'(\tilde R)-2f(\tilde R)=0$, we can solve for $b$
\be
b_\pm=-1+\mu\pm\sqrt{\mu(\mu-2n)}.
\ee
As the action must be real, $b$ must have a real value as well. This leads to a constraint $\mu>2n$. The constant $R_c$ is a free scaling parameter and for simplicity we have chosen $R_c=\tilde R$. The bound \eqref{fr0} translates to
\be
|f'(\tilde R)-1|=\frac{2n\mu}{(1+b_\pm)^2}<4\times 10^{-7}
\ee
For $b_-$ we have
\be
|f'(\tilde R)-1|=\frac{2n\mu}{(\mu-\sqrt{\mu(\mu-2n)})^2}=\frac{2 n}{\mu\Big(1-\sqrt{1-\frac{2n}\mu}\Big)^2}<4\times 10^{-7}.
\ee
With the condition $\mu>2n$ the square root can be expanded as a series. This results in $|f'(\tilde R)-1|\sim\mu<10^{-7}$ which is in clear contradiction with $\mu>2n$. Therefore we must choose $b=b_+$, for which we find
\be
\frac{2n\mu}{(\mu+\sqrt{\mu(\mu-2n)})^2}\sim\frac{2 n\mu}{4\mu^2}=\frac n{2\mu}<4\times 10^{-7} 
\ee
when $\mu>>1$. This further translates to $\mu>10^6$. Here we have assumed $n\sim 1$. For viable models this is a reasonable assumption \cite{DeFelice:2010aj}. In any case the maximum effect of $n$ is one magnitude for viable models. As $\mu>>1$ we can write the square of the graviton mass as a series of $x=1/\mu$
\be
m_g^2=\tilde R\Big(\frac 2{3n(1+2n)x}-\frac{2n(5+2n)}{3(1+2n)^2}x\Big)+\mathcal O(x^2).
\ee
Therefore, we have $n m_g^2/\tilde R\sim\mu$. As the gravitational wave observations set an upper limit for the graviton mass we find a upper limit for $\mu$ as well. We can write the relation of the background curvature to the cosmological constant as $\tilde R=4\Lambda$. Using the density parameter $\Omega_\Lambda$ we can also write
\be
\Lambda=3 H_0^2\Omega_\Lambda,
\ee
where $H_0$ is the Hubble parameter. Using the Planck collaboration results \cite{Ade:2015xua} and the LIGO results, we can now constrain the parameter $\mu$ in Hu-Sawicki models (again assuming $n\sim 1$)
%In natural units H_0\sim 10^{-42}GeV
\be
10^{20}>\mu>10^6.
\ee

We can see now, that the model is contained to a certain bracket, which is yet too constraining. However, with further more accurate measurements it is possible to further narrow down the bracket. As the gravitational wave constraints are independent of {\it e.g.} solar constraints, these offer valuable proof to the limits of $f(R)$ and scalar tensor gravity as well.

It is also interesting to notice, that the galaxy cluster limit for the graviton mass is 7 orders of magnitude tighter than the LIGO limit. If we could apply this limit, the upper limit would be of the same order as the lower limit, causing severe fine-tuning issues. However, we stress that the model dependent galaxy cluster result cannot be used directly with $f(R)$ theories.

Similar procedures can be subjected to other $f(R)$ models as, such as the Starobinsky model \cite{Starobinsky2007}, which is described by
\be
f(R)=R+\lambda R_0\Big(\big(1+\frac{R^2}{R_0^2}\big)^{-n}-1\Big).
\end{equation}
with $\lambda$ and $R_0$ positive constants and $n\in\mathbb N$. For the Starobinsky model, we can follow similar procedures to find $10^{-20}<\lambda<10^{-8}$ with the similar assumption $n\sim 1$. In a similar manner constraints could be found on any other viable model as well.

\section{Discussion}
We have studied $f(R)$ theories and the naturally emerging massive graviton. With bounds on the graviton mass produced by the gravitational wave observations it is possible to constrain $f(R)$ theories. As a case study, we concentrate on the Hu-Sawicki model. For this model we find an upper limit for the free parameter in addition to the lower limit previously presented in the literature. While the free parameter bracket is still wide, it tells a story of fine-tuning. As the massive graviton is characteristic of $f(R)$ theories and massive Brans-Dicke theories, the viability of these models is more and more under question.

The same procedure can be subjected to other $f(R)$ theories as well. As there is a known connection between $f(R)$ gravity and scalar-tensor gravity ({\it e.g.} \cite{Sotiriou:2006hs}), these theories are also a possible target for application.\footnote{The equivalent Brans-Dicke theories have a massive graviton as well. However, this is not the case of all Brans-Dicke theories.}.

The LIGO measurement accuracy is expected to rise in the future \cite{PhysRevLett.116.061102, lrr-2016-1} with the construction of additional measuring stations. As these are likely to bring down the upper limit for the graviton mass, the bracket found for the free parameter for the Hu-Sawicki model is bound to narrow down even further.

Space-based detection of gravitational waves in the future with eLISA or similar programs are expected to give constraints on the graviton mass \cite{AmaroSeoane:2012je,AmaroSeoane:2012km, Berti:2011jz}. Single observations with the space-based devices are expected to reach a two magnitudes more precise measurement than the LIGO. However, as the there are multiple events during the mission, the total accuracy is expected to be 3 orders magnitude better. This will lead to a considerably tighter bracket for viable $f(R)$ models.

Related to these limits, besides other things, detection of a non-zero graviton mass would have far-reaching consequences for $f(R)$ theories and naturally GR itself. As the $f(R)$ models predict a massive graviton, the detected mass would further fine-tune the possible parameter space. On the other hand it would spell disaster for standard GR and emphasize the need for modified gravity.

Another possibility would be to use the so-far model dependent graviton mass constraints from galaxy clusters. In order to achieve this, the effects of modified gravity on dynamics and dark matter assumptions have to be carefully considered. As these model dependent limit a far tighter than the LIGO limits, they could provide far more stringent constraints and even rule out theories considered viable.

%\section*{References}
%\bibliographystyle{iopart-num}
%\bibliographystyle{apj}
\bibliographystyle{apsrev4-1}
\bibliography{refs}

\end{document}